\documentclass[11pt]{article}

\usepackage{amsmath,amssymb,amsfonts}
\usepackage{graphicx}
\usepackage{color}
\usepackage{fullpage}
\usepackage{cite}

\def\BibTeX{{\rm B\kern-.05em{\sc i\kern-.025em b}\kern-.08em
		T\kern-.1667em\lower.7ex\hbox{E}\kern-.125emX}}

\def\fmaxA{f_{\rm maxA}}

\def\Amax{A_{\rm max}}
\def\Nr{N_{\rm r}}
\def\eps{\varepsilon}
\def\ko{k_{\scriptstyle o}}
\def\epso{\eps_{\scriptstyle 0}}
\def\muo{\mu_{_{\rm 0}}}
\def\Lm{\L_{\rm m}}
\def\##1{{\bf #1}}
\def\=#1{\underline{\underline #1}}
\def\E0{{E}_{0}}
\def\E0inc{{E}_{0}^{inc}}

\def\##1{{\bf #1}}
\def\=#1{\underline{\underline #1}}

\def\eps{\varepsilon}
\def\epso{\eps_{\scriptstyle 0}}
\def\muo{\mu_{\scriptstyle 0}}
\def\ko{k_{\scriptstyle 0}}

\def\lambdao{\lambda_{\scriptstyle 0}}

\def\sigmagr{\sigma_{\rm gr}}
\def\taugr{\tau_{\rm gr}}
\def\muyig{\=\mu_{\rm YIG}}
\def\epsyig{\eps_{\rm YIG}}

\def\Nr{N_{\rm r}}

\def\Edc{E_{\rm 0}}
\def\vEdc{\#E_{\rm 0}}

\def\Hdc{H_{\rm 0}}
\def\vHdc{\#H_{\rm 0}}

\def\Lsub{L_{\rm sub}}
\def\Lm{L_{\rm m}}

\def\LPVC{L_{\rm PVC}}

\def\epsPVC{\eps_{\rm PVC}}

\def\qe{q_{\rm e}}
\def\kB{k_{\rm B}}
\def\muc{\mu_{\rm c}}

\def\Vmum{~V~$ \mu $m$ ^{-1} $}

\def\MHzVmum{~MHz~V$^{-1}$~$\mu$m}

\def\kAm{~kA~m$^{-1}$}

\def\MHzkAm{~MHz~kA$^{-1}$~m}

\def\fmaxA{f_{\rm maxA}}
\def\Amax{A_{\rm max}}

\def\rco{\rho_{\rm xx}}
\def\rcr{\rho_{\rm yx}}

\def\bw{\triangle f _{\rm A\geq 0.9}}

\def\ux{\hat{\#x}}
\def\uy{\hat{\#y}}
\def\uz{\hat{\#z}}

\begin{document}

\begin{center}

\textsc{Design of X-Band Bicontrollable Metasurface Absorber Comprising Graphene Pixels on  Copper-Backed YIG Substrate}\\

\vspace{0.5cm}

{Govindam Sharma$^1$}, {Akhlesh Lakhtakia$^2$}, and {Pradip Kumar Jain$^1$}\\

\vspace{0.5cm}

$^1${Department of Electronics and Communication Engineering}, {National Institute Technology Patna}, 
	{Patna} {800005}, {India}\\
$^2${Departtment of Engineering Science and Mechanics}, {The Pennsylvania State University}, 
           {University Park},
           {PA 16802}, {USA}
        
\end{center}
      
\vspace{0.5cm}

\begin{abstract}
The planewave response of
a bicontrollable  metasurface absorber with graphene-patched pixels was simulated in the X band using commercial software. Each square meta-atom is a 4$\times$4 array of 16 pixels, some patched with graphene and the others unpatched.  The pixels are arranged on a PVC skin which is placed on a copper-backed YIG substrate. Graphene provides electrostatic controllability
and YIG provides magnetostatic controllability. Our design delivers  absorptance $\geq 0.9$ over a $ 100 $-MHz 
spectral regime in the X band, with $360$~\MHzkAm~magnetostatic controllabity rate and $1$~\MHzVmum~electrostatic controllability rate. Notably, electrostatic control \textit{via} graphene in the GHz range is novel.   \\ 

\end{abstract}

\textbf {Keywords:}
Bicontrollability, Magnetostatic controllability, Electrostatic controllability, Pixelation, Graphene, Yttrium iron garnet, Meta-atom, Metasurface, GHz.

\section{Introduction}
\label{sec:Intro}
Metasurfaces are thin compared to the operational wavelength, accounting for their popularity in the R\&D arena. 
The use of  materials that respond electromagnetically to a stimulus allows  controllable metasurfaces to be designed for
beam-steering reflectors/filters~\cite{PCWu}, mirrors/lenses with  variable focus~\cite{PDing}, and  
absorbers/filters~\cite{kuP,kuP1} in a wide spectrum beginning with the microwave frequencies and ending with the
visible frequencies.

Typically, controllable metasurfaces are designed to operate at high frequencies~\cite{QHe}. Direct scaling \cite{Sinclair, Lakh-scale} of controllable metasurface absorbers from THz frequencies to GHz frequencies is not always feasible, since constitutive parameters are frequency dependent. Generally, at GHz frequencies, metal is used to design the top layer of the metasurface, but using materials such as
ferrites~\cite{YHuang}, graphene~\cite{DYi}, and conductive rubber~\cite{KQiu} allow control of
metasurface absorbers.

Huang \textit{et al.} experimentally demonstrated a magnetostatically controllable
(or tunable) X-band absorber containing a ferrite slab, with a 300-MHz controllability range for  absorptance $A>0.9$~\cite{YHuang}.   Fallahi \textit{et al.} 
design an electrostatically controllable metasurface absorber containing patterned graphene---but only the maximum absorptance $\Amax$, not the
maximum-absorptance frequency $\fmaxA$,  can be controlled with that design~\cite{FallA}.  Yi \textit{et al.} used shape memory polymers to thermally control $\fmaxA \in[11.3,13.5]$~GHz~\cite{JYi}. None of these metasurface absorbers covers the complete X band with  absorptance in excess of $0.9$, which is an important requirement for wide use.  

Bicontrollable X-band metasurface absorbers are desirable for weather radar, police speed radar, and direct broadcast
television. With that in mind,
Sharma \textit{et al.} designed a pixellated  metasurface absorber 
with coarse magnetostatic and fine thermal controllability of $\fmaxA$ over the entire X band\\~\cite{GSharma}. 
The meta-atoms in this design comprise yttrium iron
garnet (YIG)-patched pixels and conductive rubber (CR)-patched
pixels on a metal-backed silicon substrate. 

Continuing in the same vein, we are now reporting 
a pixelated metasurface absorber with $\fmaxA$ controllable  both magnetostatically and electrostatically
in the entire X band,
while keeping $A\geq0.9$ over a $ 100 $-MHz spectral regime. Each  meta-atom is a $\Nr\times\Nr$ array of  pixels, some patched with graphene and the others unpatched. In contrast to numerous designs \cite{RKMaha,AYou},
the patches are not metallic. The pixels are arranged on a PVC skin which is placed on
top of a copper-backed YIG substrate. Graphene provides electrostatic controllability
and YIG provides magnetostatic controllability. Pixel size as well as the configuration of patched pixels were decided by examining the absorptance spectrums of many designs.

The plan of this paper is as follows. Section~\ref{sec:STR} provides information  on the metasurface geometry, the relative permeability dyadic of YIG,  the surface conductivity of graphene, and  theoretical simulations. Numerical results are presented and discussed in Sec.~\ref{sec:NUM}. Some remarks in Sec.~\ref{sec:CON} conclude the paper.

An $\exp(j\omega t)$ dependence on time $t$ is implicit, with $j=\sqrt{-1}$, $\omega = 2\pi f$ as the angular frequency, and $ f $ as the linear frequency. The free-space wavenumber is denoted by $ k_0 $ = $ \omega \sqrt{\epso \muo} =2\pi/\lambdao$, 
where $\lambdao$ is the free-space wavelength,
$ \epso $ is the free-space permittivity, and $\muo$ is the free-space permeability. Vectors are denoted by boldface letters; the Cartesian unit vectors are denoted by $ \ux $, $ \uy $, and $ \uz $; and dyadics are double underlined.

\section{Materials and Methods} \label{sec:STR}

\subsection{Device Structure}\label{sec2.1}
\begin{figure}[h]
	\begin{center}
		\includegraphics[width=3.5in]{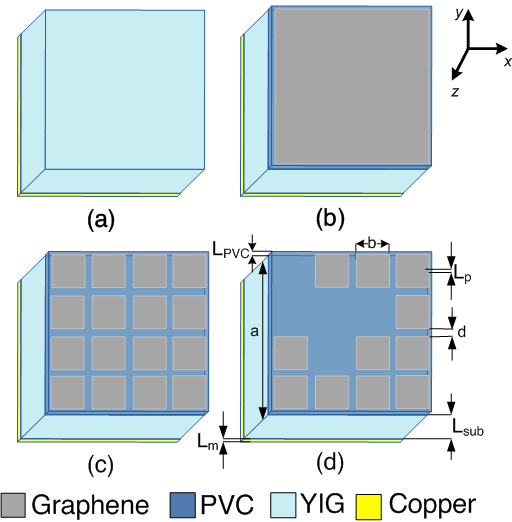}
		\caption{Schematics of four meta-atoms: (a) copper-backed   YIG substrate;
			(b) graphene   on top of a PVC skin overlaying a copper-backed   YIG substrate; (c) the
			same as (b) but with graphene   partitioned as a $4\times4$ array of graphene
			patches; and (d) the same as (c) but with only ten pixels  patched with graphene.  The Cartesian coordinate system is also shown.  
		}
		\label{fig:Str}
	\end{center}
\end{figure}

The metasurface extends to infinity in all directions in the $xy$ plane, but it is of finite thickness along the $z$ axis,
as depicted in Fig.~\ref{fig:Str}. The metasurface is a biperiodic array of square meta-atoms whose sides
are aligned along the $x$ and $y$ axes. 

Each meta-atom is of side $a$. The front surface of the meta-atom
is an array of $\Nr \times\Nr$ square pixels of side $b$, each pixel separated from every
neighboring pixel by a distance $d {\ll} a$, so that  $\Nr = a/(b + d)$. Some of the pixels
are patched with graphene, but others are not. Underneath the pixels
is a  polyvinyl-chloride (PVC) skin of thickness $\LPVC$, a YIG substrate of thickness
$\Lsub$, and a copper sheet of thickness $\Lm$ serving as a back reflector.

We fixed $\Nr=4$, $\LPVC=0.08$~mm, $\Lsub=0.2$~mm, and $\Lm=0.07$~mm. In addition, we fixed
$a  = 6$~mm, $b  = 1.45$~mm, and $d=0.05$~mm,
after multiple iterations of parameter sweeps.  

\subsection{YIG}
The relative permeability dyadic $ \muyig $  of YIG  depends on the magnitude and direction of the external magnetostatic field $\vHdc$. With this field aligned along the $x$ axis (i.e, $\vHdc=\Hdc\ux$), we have~\cite{PozarDM}
\begin{subequations}
	\begin{equation}\label{eq1}
		\muyig	
		=\ux\ux+\mu_{\rm yy}\left(\uy\uy+\uz\uz\right)+j\mu_{\rm yz}\left(\uy\uz-\uz\uy\right)\,,
	\end{equation}
	where
	\begin{equation}\label{eq2}
		\mu_{\rm yy}=  1+  
		\frac{4\pi \muo^2\gamma M_{\rm s}\left(\Hdc+j\frac{\triangle  H}{2}\right)}{\left(\muo \gamma\right)^2\left(\Hdc+j \frac{\triangle  H}{2} \right)^2-\omega^2 }\,
	\end{equation}
	and
	\begin{equation}\label{eq3}
		\mu_{\rm yz}= \frac{  4\pi \omega\muo \gamma M_s}{\left(\muo \gamma\right)^2\left(\Hdc+j\frac{\triangle H}{2}\right)^2-\omega^2}\,. 
	\end{equation}
\end{subequations}
In these equations,  
$\gamma=1.76 \times 10^{11}$~C~kg$^{-1} $ is the gyromagnetic ratio,
$ \triangle  H = 1.98$~kA~m$^{-1} $ is the resonance linewidth, and
$ M_{\rm s}  = 0.18$~Wb~m$^{-2}$ is the saturation magnetization.
The relative permittivity scalar of YIG is $ \epsyig =15$.  Note that $\Hdc\ux$ can be applied  by placing the metasurface between two magnets, so long as the lateral extent of the metasurface is in excess of $10\lambdao$.

\subsection{Graphene}
Graphene is not affected significantly by $\Hdc\ux$, because
that   magnetostatic field is wholly aligned in the plane containing the carbon atoms
\cite{Hanson}. It is, however, affected by the external electrostatic field $\vEdc=\Edc\uz$ aligned
normal to that plane, which can be applied using transparent electrodes significantly above and below
the metasurface.

The surface conductivity of graphene $\sigmagr$ comprises an intraband term and an interband term,
the latter being negligibly small compared to the former in the X band \cite{Hanson,MYGeng}. Accordingly,
\begin{eqnarray}
	\nonumber
	&&
	\sigmagr = -j\frac{\qe^2 \kB T}{\pi \, \hbar^2(\omega - 2j\taugr^{-1})}\times 
	\\[5pt]
	&&
	\left\{\frac{\muc}{\kB T} + 2\ln
	\left[1+\exp\left({-\frac{\muc}{\kB T}}\right)
	\right]
	\right\}\,,
\end{eqnarray}
where $\qe = 1.602~177 \times 10^{-19}$~C is the elementary charge, 
$ \kB=1.380~649 \times 10^ {-23}$~J~K$^{-1}$ is the Boltzmann constant, and
$ \hbar   = 1.054~572 \times 10^{-34} $~J~s is the reduced Planck constant. 
All calculations were made
for  temperature $T=300$~K.
We fixed
the momentum relaxation time $\taugr = 0.4$~ps after examining  values of the
maximum absorptance $\Amax$ and
the controllability rate $\partial \fmaxA/\partial \Edc$   for $\taugr\in[0.01,1]$~ps.
This
relaxation time can be controlled by impurity level \cite{kuP}. 

The value of the chemical potential $\muc$ depends on $\Edc$
as well as on the d.c. relative permittivity $\epsPVC = 2.7$ of PVC~\cite{Riddle}.  Thus \cite{Hanson,kuP1},
\begin{eqnarray}
	\nonumber
	&&\frac{\pi\epso\hbar^2\upsilon_{\rm F}^2}{\qe\kB^2T^2}\epsPVC\,\Edc
	=
	{\rm Li}_2\left[-\exp\left(-\frac{\muc}{\kB T}\right)\right]
	\\[5pt]
	&&\qquad-
	{\rm Li}_2\left[-\exp\left(\frac{\muc}{\kB T}\right)\right]\,,
	\label{muc}
\end{eqnarray}
where $\upsilon_{\rm F} = 10^ 6 $~m s$^{-1}$~\cite{KSNovo} is the Fermi speed for graphene and
${\rm Li}_\nu(\zeta)$ is the polylogarithm function of order $\nu$ and argument $\zeta$ \cite{poly}.
The Newton--Raphson technique \cite{Jaluria} was used to determine $\muc$ as a function
of $\Edc$.

\subsection{Theoretical Simulations}

The pixels of the  metasurface were taken to be illuminated by  a normally incident, linearly polarized 
plane wave
whose electric field phasor can be written as
\begin{equation}\label{eq4}
	\textbf{E}^{\rm inc}  = \alpha~\ux~\exp(-j \ko z),
\end{equation}
with $\alpha$ as its amplitude. 

As the metasurface is periodic along the $x$ and $y$ axes, the reflected field must be written as a doubly infinite series of Floquet harmonics ~\cite{FAhm}. Since $a<\lambdao/4$ in the entire X band,
only specular components of the reflected field are non-evanescent. Therefore, the reflected electric field 
as $z\to-\infty$  may be written as
\begin{equation}\label{eq5}
	\textbf{E}^{\rm ref} \simeq \alpha \left(\rco \ux +\rcr\uy\right)
	\exp(j \ko z),
\end{equation}
where $\rco\in\mathbb{C}$ is the co-polarized reflection coefficient and $\rcr\in\mathbb{C}$ is the cross-polarized reflection coefficient. The transmitted field in the region beyond the metallic back reflector was negligibly small in magnitude, because
$\Lm$ is much larger than the 
penetration depth in copper. Hence, the
absorptance was calculated as
\begin{equation}\label{eq6}
	A=1-(\vert\rco\vert^2+\vert\rcr\vert^2). 
\end{equation}

Normal incidence on several configurations of the pixelated-metasurface 
absorber was simulated using the commercial tool CST Microwave Studio\texttrademark~2020. Periodic boundary conditions were applied along the $x$ and $y$ axes. The
option \textsc{open} was chosen
for the $z$ axis  and the \textsc{planewave} condition  applied.
The meta-atom was partitioned into as many as 10,026 tetrahedrons for each simulation in order to achieve  convergent results. The 
absorptance $ A $ was calculated for  $f~\in~[8, 12]$~GHz,   $  \Hdc  \in [180, 270]$~\kAm, and  $ \Edc \in [0, 100]$~\Vmum.

\section{Numerical Results} \label{sec:NUM}

\begin{figure}[h]
	\begin{center}
		\includegraphics[width=3.5in]{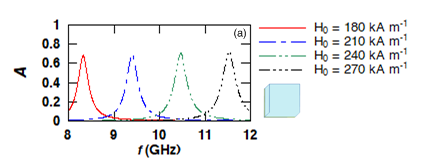}
		\includegraphics[width=3.5in]{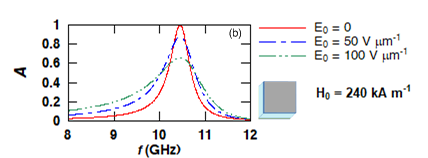}
		\caption{Absorptance spectrums  of (a) the YIG/copper structure of Fig.~\ref{fig:Str}(a) for  $\Hdc\in\{180, 210, 240, 270\}$~\kAm and (b) the graphene/PVC/YIG/copper structure of
			Fig.~\ref{fig:Str}(b) for $\Edc \in \{0, 50, 100\}$~\Vmum and
			$\Hdc = 240$~\kAm.
		}
		\label{fig:All_meta}
	\end{center}
\end{figure}

We begin by discussing the response of the copper-backed YIG substrate shown in Fig.~\ref{fig:Str}(a).
Figure~\ref{fig:All_meta}(a) shows the computed spectrums of $ A $ for $ \Hdc \in \{180, 210, 240, 270\}$~\kAm, this metasurface being unaffected by $\Edc$. The maximum-absorptance frequency $ \fmaxA $ blueshifts
as the magnetostatic field $ \Hdc $ increases, but the maximum absorptance $ \Amax \leq 0.8$. Hence, 
the copper-backed YIG substrate does not satisfy the  requirement of
$\Amax \in[0.9,1]$ in any spectral regime within the X band.

Covering the YIG substrate on the top, first by a PVC skin
and then by graphene, as in   Fig.~\ref{fig:Str}(b), certainly affects the absorptance. Graphene
makes this structure susceptible to $\Edc$, in addition to the YIG-mediated susceptibility to $\Hdc$.
The  spectrums of $A$ are shown in Fig.~\ref{fig:All_meta}(b) for \sloppy{ $\Edc \in \{0, 50, 100\} $}~\Vmum~and
$\Hdc = 240$~\kAm. Now, $\Amax$ becomes a decreasing function of $\Edc$, although the
controllability of $\fmaxA$ by $\Hdc$ (results not shown) is maintained. Therefore, the copper-backed YIG substrate
with or without the  graphene/PVC bilayer is inadequate as the desired bicontrollable metasurface absorber.

\begin{figure}[h]
	\begin{center}
		\includegraphics[width=3.5in]{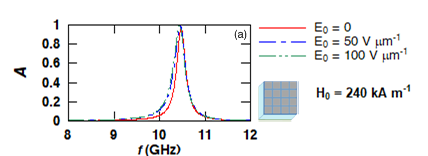}
		\includegraphics[width=3.5in]{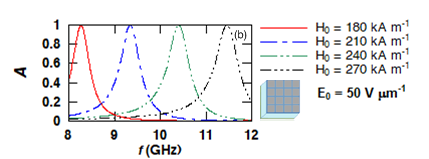}
		\caption{Absorptance spectrums  of  the pixelated metasurface  of
			Fig.~\ref{fig:Str}(c), with all 16 pixels per meta-atom patched with graphene. (a) $\Edc \in \{0, 50, 100\}$~\Vmum~and
			$\Hdc = 240$~\kAm. (b) $\Hdc\in\{180, 210, 240, 270\}$~\kAm~and $\Edc=50$~\Vmum.
		}
		\label{fig:allpix}
	\end{center}
\end{figure}
\begin{figure}[h]
	\begin{center}
		\includegraphics[width=3.5in]{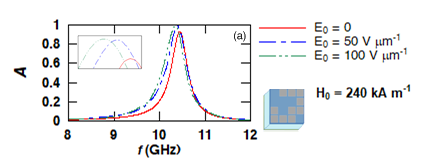}
		\includegraphics[width=3.5in]{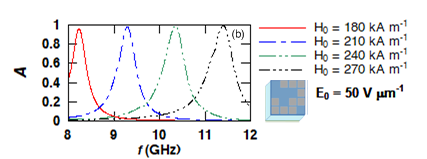}
		
		\caption{Absorptance spectrums  of  the pixelated metasurface  of
			Fig.~\ref{fig:Str}(d), with only 10 pixels per meta-atom patched with graphene. (a) $\Edc \in \{0, 50, 100\}$~\Vmum~and
			$\Hdc = 240$~\kAm. (b) $\Hdc\in\{180, 210, 240, 270\}$~\kAm~and $\Edc=50$~\Vmum.
		}
		\label{fig:leftpix}
	\end{center}
\end{figure}

For the next set of simulations, we partitioned the graphene in Fig.~\ref{fig:Str}(b) into 16 patches per meta-atom, as shown
in   Fig.~\ref{fig:Str}(c). The absorptance spectrums in
Fig.~\ref{fig:allpix}(a) for \sloppy 
{$\Edc \in \{0, 50, 100\}$}~\Vmum and
$\Hdc = 240$~\kAm~clearly indicate that   pixelation can increase $\Amax$ and make it less
susceptible to variations in $\Edc$, when compared with the spectrums in Fig.~\ref{fig:All_meta}(b).
The absorptance spectrums in
Fig.~\ref{fig:allpix}(b) for $\Hdc\in\{180, 210, 240, 270\}$~\kAm~and $\Edc=50$~\Vmum~confirm the magnetostatic
controllability of $\fmaxA$.

Finally, we present the absorptance spectrums calculated for the metasurface of  Fig.~\ref{fig:Str}(d), which has ten
graphene-patched and six unpatched pixels. The specific configuration of unpatched pixels
was selected after studying the absorption spectrums for
many other configurations. The spectrums in Fig.~\ref{fig:leftpix}(a) for $\Edc \in \{0, 50, 100\}$~\Vmum~and
$\Hdc = 240$~\kAm~and Fig.~\ref{fig:leftpix}(b) for $\Hdc\in\{180, 210, 240, 270\}$~\kAm~and $\Edc=50$~\Vmum~
indicate that a bicontrollable spectral regime with $A\geq0.9$ and $\Amax\approx0.99$  
can be achieved with $360$~\MHzkAm~magnetostatic control and $1$~\MHzVmum~electrostatic control
of $\fmaxA$. 
Coarse control is possible through $\Hdc$ and fine control through $\Edc$.
The bandwidth $\bw$ of this absorber is about $ 100 $~MHz, which is suitable for many X-band applications.

Table~\ref{Tab:Compare} compares the proposed metasurface absorber with previously reported absorbers. 
Yuan \textit{et al.}~\cite{YuanH} designed a voltage-controlled  metasurface 
absorber containing   varactor diodes, for   X-band
operation  with   $\fmaxA$ controlled in a 440-MHz range. 
Huang \textit{et al.}~\cite{YHuang} incorporated a   meta-atom with a metal resonator printed on FR4 and affixed to a metal-backed ferrite substrate. Their metasurface absorber has a wider bandwidth than the proposed absorber
red{has}, but the controllability range is smaller than of the proposed absorber. Sharma \textit{et al.}~\cite{GSharma} reported a  meta-atom with a square array of pixels patched with conductive rubber and YIG on a metal-backed silicon substrate. This bicontrollable metasurface has a wider bandwidth with stable maximum absorptance in the entire X band, and fine control is thermal rather than electrostatic as for the proposed absorber.

I 
	\begin{table}[h]
		\caption{Structure, type of control, controllability range of maximum-absorptance frequency ($\fmaxA$), bandwidth ($ \bw $), and controllability rate of reported metasurface absorbers and the proposed metasurface absorber.}
		\label{Tab:Compare}
		\begin{center}
			\begin{tabular}{|c|c|c|c|c|c|}
				\hline
				Ref. & Structure & Control & $\fmaxA$ &  $\bw$ & Controllability \\
				& & method(s)&(GHz)& (MHz)& rate  \\
				\hline
				\citenum{YHuang}& Metal resonator/FR4/ &    {magnetostatic}  & 9.3--9.7 & 150 &   3~\MHzkAm  \\
				&ferrite/metal sheet & & & & \\
				
				\hline
				\citenum{YuanH} & Metal pads separated by & &	& &\\
				& varactor diodes/FR4 &   {electrical} & 8.25--9.25 & 400 & 100~MHz~V$^{-1}$  \\
				& sheet/metal sheet& &	& &\\
				
				\hline
				\citenum{GSharma} & YIG-  and CR-patched pixels/   &   {magnetostatic}   & 8--13 & 200 &360~\MHzkAm   \\
				& silicon/metal sheet & and {thermal} & & & and 1~MHz~K$^{-1}$  \\
				
				\hline
				
				This 		   & Graphene pixels/PVC & {magnetostatic}   & 8--12  &  100 & 360 \MHzkAm \\
				
				work	       & skin/YIG/metal sheet & and {electrostatic}	 & 	& &and $1$~\MHzVmum\\
				\hline
			\end{tabular}
		\end{center}
	\end{table}

\section{Concluding Remarks} \label{sec:CON}

We conceived, designed, and investigated  a electrostatically and magnetostatically
controllable  metasurface absorber for operation in the entire X band. The meta-atom comprises  ten graphene-patched pixels and six unpatched pixels in a 4$\times$4 array on a PVC skin   that is affixed to a metal-backed YIG substrate. Graphene provides electrostatic controllability and YIG provides magnetostatic controllability. Electrostatic control of the maximum-absorptance frequency using  graphene-patched pixels in the GHz range is novel. The configuration
of graphene-patched and unpatched pixels was optimized to achieve stable maximum absorptance of $0.99$, with pixelation performing better than continuous graphene. According to our simulations, the chosen design delivers  absorptance $\geq 0.9$ over a $ 100 $-MHz band, with $360$~\MHzkAm~magnetostatic controllabity rate and $1$~\MHzVmum~electrostatic controllability rate.  The proposed X-band absorber can be used to improve the performance of   radar systems.

\end{document}